\documentclass[12pt,a4paper]{article}
\usepackage[english]{babel}
\usepackage{amsmath,amsfonts,amssymb,amsthm}
\newcommand{\cK}{\mathcal{K}}
\newcommand{\cS}{\mathcal{S}}
\newcommand{\cH}{\mathcal{H}}

\newcommand{\cE}{\mathcal{E}}
\newcommand{\cI}{\mathcal{I}}
\numberwithin{equation}{section}
\theoremstyle{plain}
\newtheorem{theorem}{Theorem}
\newtheorem{proposition}[theorem]{Proposition}
\newtheorem{corollary}[theorem]{Corollary}

\begin{document}
\title{
Independent electrons model for open quantum systems:
Landauer-B\"uttiker formula and strict positivity of the entropy production}
\author{Gheorghe Nenciu\\ 
Dept.  Theor. Phys., Univ. of Bucharest\\ P.O. Box MG 11, RO-077125, 
Bucharest, Romania\\
and \\ 
Institute of Mathematics of the Romanian Academy\\  
PO Box 1-764, 
RO-014700 Bucharest, Romania \\
E-mail Gheorghe.Nenciu@imar.ro}
\date{}
\maketitle

\begin{abstract}

A  general argument leading from the formula for currents
through an open noninteracting mesoscopic system given by the theory of non-equilibrium
steady states (NESS) to  the Landauer-B\"uttiker formula is pointed out. Time reversal symmetry is not assumed. As a consequence it follows that, as far as the system has a nontrivial scattering theory and the reservoirs have different temperatures and/or chemical potentials, the entropy production is strictly positive.
\end{abstract}

\newpage

\section{Introduction}
 
Landauer-B\"uttiker type formulas i.e. expressions relating the
(charge, energy etc) currents through mesoscopic systems connected
with electron reservoirs to the corresponding transmission
coefficients have been proved to be a key tool for
analyzing the quantum conductance in nanostructures. Obtained initially
for the stationary case by phenomenological arguments \cite {BILP},
\cite{B}, \cite{IL} they have been widely extended and used. As for the
derivation, one usually assumes that the reservoirs have a lead
geometry and in order to make use of the asymptotic form of the
scattered state the current is evaluated far away from the scatterer
\cite{BS}, \cite{FL}, \cite{MPD}, a
procedure justified (at least in the stationary regime) by charge
conservation. However this approach may become problematic for other
reservoir geometries when the leads are short or even inexistent (see
e.g. \cite{HDQEB}, \cite{BBMH}) or in non-stationary regime.

At a more basic level one starts from a non-equilibrium statistical
mechanics formulation (e.g. linear response theory, NESS theory etc)
and the problem of proving the Landauer-B\"uttiker formula is to show
that the obtained formula for the current can be cast in a form in
which the structure of the mesoscopic system enters only via its
transition matrix of the associated scattering problem as suggested by
the phenomenological derivation.
In this note we shall consider the formula for the current as given by NESS theory (see e.g \cite{Ru1}, \cite{FMU}, \cite{AJPP}, \cite{JKP} and references 
therein).
 One starts at $t=0$ from an equilibrium state of
the {\em decoupled} system (i.e. no coupling between the mesoscopic
system and the reservoirs) with reservoirs having different
temperatures and/or chemical potentials. At the one particle level the
system is described by
\begin{equation}\label{decmodel}
H_{0}=H_{\cS}+\sum_{j=1}^{N}H_{j};\cH= \cH_{\cS}\oplus_{j=1}^{N}\cH_{j}
\end{equation}
 where $\cH_{\cS},\;
H_{\cS}$ are the Hilbert space and hamiltonian of the mesoscopic
system and $\cH_{j},\;H_{j}$ are the Hilbert space and the hamiltonian
of the jth reservoir. The coupling, described at the one particle
level by $V$, is switched on suddenly at $t=0$. In the limit
$t\rightarrow \infty$ the system settles down to a non-equilibrium stationary state. The currents out from the reservoirs are defined as minus
the time
variation of their charge. Since the electrons are considered
independent the second quantization machinery allows to write the
currents as given by the general NESS theory  in terms of one-particle objects. More precisely
if $\beta_{j},\;\mu_{j}$ are the temperature and the chemical
potential respectively of the jth reservoir in the initial state then
the current out from the kth reservoir in the ``final'' steady state
is \cite{AJPP} (the charge of the electron is $-e$):
\begin{equation}\label{curent}
j_{k}=ieTr_{\cH}(\Omega_{+}\Pi_{0}F_{0}\Pi_{0}\Omega_{+}^{*}[V,\Pi_{k}])
=ieTr_{\cH}(\Pi_{0}F_{0}\Pi_{0}\Omega_{+}^{*}[V,\Pi_{k}]\Omega_{+})
\end{equation}
where $\Pi_{j}$ are the orthogonal projections onto $\cH_{j}$ in
$\cH$, $\Pi_{0}=\sum_{j=1}^{N}\Pi_{j}$,
\begin{equation}\label{moller}
\Omega_{
+}=s-\lim_{t\rightarrow
 - \infty}e^{it(H_{0}+V)}e^{-itH_{0}}\Pi_{0}
\end{equation}
 (we follow the notation  in the physical literature and \cite{RS}) and
\begin{equation}\label{initial}
\Pi_{0} F_{0}\Pi_{0}=\sum_{j=1}^{N}f_{\beta_{j},\mu_{j}}^{FD}(H_{j}).
\end{equation}
Here $f_{\beta_{j},\mu_{j}}^{FD}(x)$ are the usual Fermi-Dirac functions:
\begin{equation}
f_{\beta_{j},\mu_{j}}^{FD}(x)= (1+e^{\beta_{j}(x-\mu_{j})})^{-1}.
\end{equation}
 Two remarks are in order here. Firstly, we would like to stress the fact that (\ref{curent}) gives {\em only} the steady part of the current; a (quasi) periodic component might also exists if $H=H_{0}+V$ has bound states (see e.g. \cite{JKP} Section 5.5 for a discussion about this point). Secondly, due to the fact that there is no interaction between electrons, (\ref{curent}) can be obtained by means of elementary second quantization theory without the use of the heavy artilery of NESS theory.

In the framework of NESS theory and under appropriate technical conditions the case $dim\;\cH_{\cS} =1$ (single state quantum dot) has been thoroughly studied in \cite{AJPP},\cite{JKP}; in particular the Landauer-B\"uttiker formula has been proved and strict positivity of the entropy production established. Actually in this case the model is nothing but the well known exactly solvable Wigner-Weisskopf Atom model (known also under the name of Friedrichs model) for which the M{\o}ller operator as well as the scattering matrix can be explicitly written down so that the Landauer-B\"uttiker formula can be directly verified. Moreover an extension of the analysis in \cite{AJPP}, \cite{JKP} to more general models similar to the one considered in the present note will be given (C.-A. Pillet: private communication) in \cite{AJPP1} announced in \cite{AJPP}.

An alternative way (and sometimes more satisfactory from the physical point of
view) of computing currents in non-equilibrium statistical mechanics is
to start at $t=-\infty$ with reservoirs at {\em the same} temperature and chemical
potential and with an  equilibrium state of the {\em coupled}
system and  then switch on adiabatically the bias in chemical
potential and/or temperature. Unfortunately due to the fact that in
this case the ``perturbation'' is not localized the problem is much
more difficult and it has been worked out only at the linear response
theory level. In this context the Landauer-B\"uttiker formula has been
shown to hold true at the heuristic level by Baranger and Stone
\cite{BS}
and rigorously proved for a tight-binding model for reservoirs by
Cornean, Jensen and Moldoveanu \cite{CJM}.

Coming back to the formula (\ref{curent}) the problem is that the
M{\o}ller operator, $\Omega_{+}$, involves only ``half'' of the
evolution from $-\infty$ to $\infty$ encoded in the scattering
matrix so one has to show that one can rewrite the current only in
terms of scattering matrix and the initial equilibrium state.
 In the related context of adiabatic quantum pumps theory it has been
proved in  \cite{AEGSS} that this is indeed the case  for the lead geometry
of the reservoirs.

The aim of this note (which is a revision and extension of \cite{N})  is a narrow one:  to outline a general argument
leading from (\ref{curent}) to the Landauer-B\"uttiker formula.
The
argument is entirely elementary and very general: it works for an
arbitrary geometry of the reservoirs (e.g. half spaces, semi infinite
leads with arbitrary section etc) and arbitrary mesoscopic systems of
finite size. Also we allow the reservoirs to be coupled both via the
mesoscopic system and by direct contacts \cite{FMU}.
 Actually, the only thing which  is needed is a  good stationary
scattering theory for the pair $(H_{0}, H_{0}+V)$. This is consistent
 with the generality of the phenomenological arguments
leading to Landauer-B\"uttiker formula. For the model at hand the strict positivity of the entropy (which is a central issue in NESS theory and has been established under various conditions for by far more general models see e.g.\cite{AJPP}, \cite{AS}, \cite{MMS} and references therein) follows from the Landauer-B\'uttiker formula; while for systems with time reversal symmetry this is straightforward by the argument in \cite{AJPP} in the general case the argument is a bit more involved.

The content of the note is as follows. In Section 2 we specify the model and give the needed formulas from stationary scattering theory. Section 3 contains the argument leading from the formula (\ref{curent}) to the Landauer-B\"uttiker formula.
In order not to burden the simplicity of the argument in technical and
notational details we shall give it at the formal level and in a
simple context: two reservoirs with the same simple (single channel) absolutely
continuous spectrum, $\sigma_{0} \subset [0, \infty)$,  and mesoscopic systems with a finite number of
states (i.e. $ dim\; \cH_{\cS}=M < \infty$). 
 For a mathematical
substantiation for various concrete models one has to make precise the technical conditions on
$H_{0}$ and $V$ and then check that one can apply the results  of the
rigorous stationary scattering theory as developed e.g. in \cite{AJS},
\cite{Y1}, \cite{Y2} (for stationary scattering theory at the formal
level we send the reader to \cite{GW}). In Section 4 we give some straightforward extensions as well as the argument for strict positivity of the entropy production.

\section{The model and its scattering theory}

As already said in the introduction, at the one particle level the uncoupled
system is described by
\begin{equation}\label{decmodel1}
H_{0}=H_{\cS}+\sum_{j=1}^{N}H_{j};\cH= \cH_{\cS}\oplus_{j=1}^{N}\cH_{j}
\end{equation}
 where $\cH_{\cS},\;
H_{\cS}$ are the Hilbert space and hamiltonian of the mesoscopic
system and $\cH_{j},\;H_{j}$ are the Hilbert space and the hamiltonian
of the jth reservoir.
We suppose that the spectral representation of $H_{j}$  $j=1,2$ is given in
terms of generalized eigenfunctions, $|\psi_{j,E}^{0}>$,  living in an
appropriate ``weighted'' Hilbert space,  $\cK_{j}^{*}$ (Gelfand triplets
structure: $\cK_{j} \subset \cH_{j} \subset \cK_{j}^{*}$) :
\begin{equation}
H_{j}|\psi_{j,E}^{0}>=E|\psi_{j,E}^{0}>,\; E\in \sigma(H_{j})=\sigma_{0} \subset [0,\infty).
\end{equation}
For $f\in \cH_{j}$ we denote by $f(E)$ its generalized Fourier
transform:
\begin{equation}\label{gft}
f(E)=<\psi_{j,E}^{0},f>.
\end{equation}

As concerning $V$,  we suppose to have the following structure (in the
decomposition given by (\ref{decmodel})):
\begin{equation}\label{V}
    V=\left(%
\begin{array}{ccc}
  0 & V_{\cS 1} & V_{\cS 2}\\
  V_{\cS 1}^{*} & 0 &  V_{12}\\
   V_{\cS 2}^{*}& V_{12}^{*}& 0
\end{array}%
\right).
\end{equation}
Since $\cH_{\cS}$ has finite dimension, $ V_{\cS 1}$ and $ V_{\cS 2}$
are finite rank operators. We suppose the ``direct contact'', $
V_{12}$,  to be also of finite rank. Accordingly:
\begin{equation}\label{cuplaj1}
 V_{\cS j}=\sum_{l=1}^{m_{j}\leq M}v_{jl}|s_{jl}><f_{jl}|,
\end{equation}
\begin{equation}\label{cuplaj2}
 V_{12}=\sum_{l=1}^{m < \infty}v_{l}|g_{1l}><g_{2l}|
\end{equation}
where $j=1,2$ ,  $\left\{f_{jl}\right\}_{l=1}^{m_{j}}$,
$\left\{g_{jl}\right\}_{l=1}^{m}$,
$\left\{s_{jl}\right\}_{l=1}^{m_{j}}$ are orthonormal systems in
$\cH_{j}$ and  $\cH_{\cS}$ respectively and $v_{jl},\;v_{l}>0$.

Since $V$ is of finite rank, by Kato-Kuroda-Birman theory \cite{RS},
\cite{Y1}, \cite{Y2} the M{\o}ller
operators
\begin{equation}\label{mollerpm}
\Omega_{\pm}=s-\lim_{t\rightarrow \mp \infty}e^{it(H_{0}+V)}e^{-itH_{0}}\Pi_{0}
\end{equation}
exist and are unitary from $\Pi_{0}\cH$ onto the absolutely continuous
subspace,
$\Pi_{ac}\cH$, of $H$.

We impose further conditions on $f_{jl}$, $g_{jl}$ in  order to assure that $\Omega_{\pm}$ provide spectral representations for $H$
restricted to $\Pi_{ac}\cH$ ie for all $E\in \sigma_{ac}(H)=
\cup_{j=1}^{n}\sigma(H_{j})$,
with a possible exception of a discrete set, $\cE$,  $\Omega_{\pm}$
have bounded extensions in the orthogonal sum of $\cK_{j}^{*}$ and
\begin{equation}
|\psi_{j,E}^{\pm}>=\Omega_{\pm}|\psi_{j,E}^{0}>
\end{equation}
are generalized eigenfunctions for $H$:
\begin{equation}
H|\psi_{j,E}^{\pm}>=E|\psi_{j,E}^{\pm}>.
\end{equation}
 A sufficient condition (which at the price of more technicalities can be
weakened) in the case when $H_{j}$ are discrete or continuous
Laplaceans supplemented with boundary conditions is that
$f_{jl}$, $g_{jl}$ are exponentially localized in space. This
condition also
implies that the generalized Fourier coefficients,
$f_{jl}(E)$, $g_{jl}(E)$ (see (\ref{gft}))  of $f_{jl}$, $g_{jl}$ are
smooth functions of $E$, a fact which is needed in order to apply the principal
value formula during the proof below.
The generalized eigenfunctions satisfy the Lippmann-Schwinger
\cite{GW},\cite{AJS}, \cite{Y1}, \cite{Y2}
equation:
\begin{equation}\label{LS}
|\psi_{j,E}^{\pm}>=|\psi_{j,E}^{0}>-(H_{0}-E \mp i0)^{-1}V|\psi_{j,E}^{\pm}>.
\end{equation}

Consider now the scattering operator
\begin{equation}\label{S}
S=\Omega_{-}^{*}\Omega_{+}
\end{equation}
and the corresponding transition operator, $T$, defined by
\begin{equation}
S=1-2i\pi T.
\end{equation}

Since $S$ (and then $T$) commutes with $H_{0}$ it has a  spectral
representation:
\begin{equation}\label{TE}
S=\int_{\sigma_{0}}S(E) dE,\;\; 
T=\int_{\sigma_{0}}T(E) dE
\end{equation}
where $S(E)$ is a unitary two by two matrix (we are considering the
case of  two
reservoirs with simple spectrum). From the unitarity of $S(E)$ it
follows that $T(E)$ satisfies the so called optical theorem:
\begin{equation}\label{toptica}
T(E)-T^{*}(E)=-2\pi i T(E)T^{*}(E).
\end{equation}
The basic result of the stationary scattering theory is the formula
for $T(E)$ in terms of the generalized eigenfunctions of $H_{0}$
\cite{GW}, \cite{AJS}, \cite{Y1}, \cite{Y2}:
\begin{equation}\label{Tpsi} 
T_{jk}(E)=<\psi_{j,E}^{0},V\psi_{k,E}^{+}>=
<\psi_{j,E}^{0},V\Omega_{+}\psi_{k,E}^{0}>.
\end{equation}

\section{Landauer-B\"uttiker formula}

To prove the Landauer-B\"uttiker formula in the context described
above amounts to prove :
\begin{proposition}

\begin{eqnarray}\label{LB}
&j_{1}=ieTr_{\cH}(\Omega_{+}\Pi_{0}F_{0}\Pi_{0}\Omega_{+}^{*}[V,\Pi_{1}])=
\nonumber
\\
&-
2e\pi\int_{\sigma_{0}} dE(f_{\beta_{1},\mu_{1}}^{FD}(E)-
f_{\beta_{2},\mu_{2}}^{FD}(E)
)|T_{12}(E)|^{2}.
\end{eqnarray}

\end{proposition}

The second equality in (\ref{LB}) is
 the main result of this note.

We compute  $j_{1}$  from (\ref{curent}). Inserting the formula for
$V$ (see (\ref{V}), (\ref{cuplaj1}), (\ref{cuplaj2})) and computing the
trace in appropriate bases one gets:
\begin{eqnarray}\label{curent1}
&j_{1}=-2e\Im (\sum_{l=1}^{l=m_{1}}v_{1l}<f_{1l},
\Omega_{+}\Pi_{0}F_{0}\Pi_{0}\Omega_{+}^{*}s_{1l}>+
\nonumber
\\
&
\sum_{l=1}^{l=m}v_{l}<g_{1l},
\Omega_{+}\Pi_{0}F_{0}\Pi_{0}\Omega_{+}^{*}g_{2l}>).
\end{eqnarray}
 Using the
spectral representation of $\Pi_{0}F_{0}\Pi_{0}$ (see (\ref{initial})) 
in (\ref{curent1}) or, alternatively, evaluating directly the trace in the
r.h.s. of (\ref{curent}) in the generalized basis of $H_{0}$ one gets:
\begin{eqnarray}\label{curent2}
& j_{1}=-
2e \int_{\sigma_{0}}
dE\{f_{\beta_{1},\mu_{1}}^{FD}(E)\Im 
<V\Omega_{+}\psi_{1,E}^{0},\Pi_{1}\Omega_{+}\psi_{1,E}^{0}>+
\nonumber
\\
&
f_{\beta_{2},\mu_{2}}^{FD}(E)\Im <V\Omega_{+}\psi_{2,E}^{0},\Pi_{1}\Omega_{+}\psi_{2,E}^{0}>
\}.
\end{eqnarray}
Let us compute first the coefficient of
$f_{\beta_{1},\mu_{1}}^{FD}(E)$ in (\ref{curent2}).
Using the Lippmann-Schwinger equation (see (\ref{LS})) for
$\Omega_{+}\psi_{1,E}^{0}$ and
 the spectral representation of $H_{0}$ one has:
\begin{eqnarray}\label{fp0}
&<V\Omega_{+}\psi_{1,E}^{0},\Pi_{1}\Omega_{+}\psi_{1,E}^{0}>=
\nonumber
\\
& 
<V\Omega_{+}\psi_{1,E}^{0},\psi_{1,E}^{0}>-
<V\Omega_{+}\psi_{1,E}^{0},\Pi_{1}\frac{1}{H_{0}-E-i0}V\Omega_{+}\psi_{1,E}^{0}>=
\nonumber
\\
& 
\overline{T_{11}(E)}-
\int dE'\frac{| <\psi_{1,E'}^{0},V\Omega_{+}\psi_{1,E}^{0}>|^{2}}{E'-E-i0}.
\end{eqnarray}
Now
the important fact is that we need only the imaginary part of
(\ref{fp0}).
Then using the principal value formula 
\begin{equation}\label{PV}
\frac{1}{x-i0}= i\pi \delta(0)+ PV\frac{1}{x},
\end{equation}
to evaluate the integral in (\ref{fp0})  ( since 
$| <\psi_{1,E'}^{0},V\Omega_{+}\psi_{1,E}^{0}>|^{2}$ depends upon $E'$
only via $f_{1l}(E')$ and  $g_{1l}(E')$ which are smooth by assumption,
this is legitimate)
 one obtains that the coefficient of
$f_{\beta_{1},\mu_{1}}^{FD}(E)$ in (\ref{curent2}) is
$$ \Im (\overline{T_{11}(E)}-i\pi |T_{11}(E)|^{2}).$$  Now  the  use of the optical
theorem (\ref{toptica}) leads to the conclusion that the coefficient of
$f_{\beta_{1},\mu_{1}}^{FD}(E)$ in (\ref{curent2}) is $\pi |T_{12}(E)|^{2}$.
A similar computation for the coefficient  of $f_{\beta_{2},\mu_{2}}^{FD}(E)$ in (\ref{curent2})
 (in this case the term linear in $T(E)$ vanishes) leads to
 $-\pi|T_{12}(E)|^{2}$ 
and the proof of (\ref{LB}) is finished.

\section{Generalizations and strict positivity of the entropy production}

We give first some straightforward extensions of the result in previous section.

i. A  similar proof applied to the energy
current (see e.g. \cite{AJPP}) gives:
\begin{eqnarray}\label{LB1}
&\Phi_{1}=-iTr_{\cH}(\Omega_{+}\Pi_{0}F_{0}\Pi_{0}\Omega_{+}^{*}[V,\Pi_{1}H_{0}\Pi_{1}])=
\nonumber
\\
&
2\pi\int_{\sigma_{0}} dE(f_{\beta_{1},\mu_{1}}^{FD}(E)-
f_{\beta_{2},\mu_{2}}^{FD}(E)
)E|T_{12}(E)|^{2}.
\end{eqnarray}

ii. 
The condition that $\sigma (H_{1})=\sigma (H_{2})$ (as  sets) is not
necessary ; in the general case only the energies in the intersection
of $\sigma
(H_{1})$ with $\sigma (H_{2})$ can have nontrivial scattering and then
contribute to the current.

iii.  The straightforward generalization of (\ref{LB}), (\ref{LB1}) to the case of
$N$ reservoirs is given by:
\begin{eqnarray}\label{LBN}
&j_{k}=ieTr_{\cH}(\Omega_{+}\Pi_{0}F_{0}\Pi_{0}\Omega_{+}^{*}[V,\Pi_{k}])=
\nonumber
\\
&-
2e\pi\int_{\sigma_{0}} dE\sum_{j=1}^{N}(f_{\beta_{k},\mu_{k}}^{FD}(E)-
f_{\beta_{j},\mu_{j}}^{FD}(E)
)|T_{
kj}(E)|^{2},
\end{eqnarray}
\begin{eqnarray}\label{LBNE}
&\Phi_{k}=-iTr_{\cH}(\Omega_{+}\Pi_{0}F_{0}\Pi_{0}\Omega_{+}^{*}[V,\Pi_{k}H_{0}\Pi_{k}])=
\nonumber
\\
&
2\pi\int_{\sigma_{0}} dE\sum_{j=1}^{N}(f_{\beta_{k},\mu_{k}}^{FD}(E)-
f_{\beta_{j},\mu_{j}}^{FD}(E)
)E|T_{
kj}(E)|^{2}.
\end{eqnarray}
Notice that for $N>2$  and  systems without time reversal symmetry one can have $|T_{
kj}(E)|^{2}\neq |T_{
jk}(E)|^{2}$.  Still $\sum_{k=1}^{N}j_{k}=0$, $\sum_{k=1}^{N}\Phi_{k}=0$ as required by charge and energy
conservation, due to the fact that the unitarity of $S$ implies that $T(E)$ is a normal matrix:
\begin{equation}\label{normal}
T(E)T^{*}(E)=T^{*}(E)T(E).
\end{equation}

iv.  If at some energy $E$ the spectra of $H_{j}$ do not have
multiplicity one then $T_{jk}(E)$  become operators
$T_{jk}(E):\cH_{j}(E) \rightarrow \cH_{k}(E)$ and $|T_{
jk}(E)|^{2}$ in (\ref{LBN})  are to be   replaced by  $Tr_{\cH_{j}(E)}T_{jk}(E)T_{jk}^{*}(E)$.

Consider now the entropy production. The entropy production rate (as given by heuristic thermodynamic arguments) has been identified as the (thermodynamic and $t\rightarrow \infty$) limit of the relative entropy of the evolved state with respect to the initial state, 
 \cite{Ru2}, \cite{JP}, \cite{FMU},\cite{AJPP}, and then its  positivity follows from Klein's inequality. However, although for finite reservoirs and times its strict positivity can be also easily obtained from Klein's inequality it is not clear whether it survives when taking thermodynamic and infinite time limits. Accordingly the strict positivity of the entropy production has to be established for each concrete realization of NESS theory and many results are known (see e.g. \cite {AS}, \cite{AJPP},\cite{MMS} and references therein). For the model considered in this note the entropy production, $\sigma$,  writes as (see e.g. formula (2.61) in \cite{FMU} or formula (6.46) in \cite{AJPP}):
\begin{equation}\label{eprp}
\sigma =-\sum_{k=1}^{N}\beta_{k}(\Phi_{k}-\mu_{k}\cI_{k})
\end{equation}
where
\begin{equation}\label{parnum}
\cI_{k}=-\frac{1}{e}j_{k}=
2\pi\int_{\sigma_{0}} dE\sum_{j=1}^{N}(f_{\beta_{k},\mu_{k}}^{FD}(E)-
f_{\beta_{j},\mu_{j}}^{FD}(E)
)|T_{
kj}(E)|^{2}
\end{equation}
is the particle current. From (\ref{epr}), (\ref{LBN}) and (\ref{parnum}) and using the notations:
\begin{equation}\label{not}
f(x)=(1+x)^{-1};\;\;\; x_{E,j}=\beta_{j}(E-\mu_{j})
\end{equation}
one obtains:
\begin{equation}\label{epr}
\sigma=-2\pi\int_{\sigma_{0}} dE\sum_{j,k=1}^{N}(f(x_{E,k})-f(x_{E,j}))x_{E,k}
|T_{k,j}(E)|^{2}.
\end{equation}
Using
\begin{equation}\label{normal1}
\sum_{m=1}^{N}|T_{l,m}(E)|^{2}=\sum_{m=1}^{N}|T_{m,l}(E)|^{2},\;\;l=1,2,...,N
\end{equation}
which follows from (\ref{normal}), one can rewrite (\ref{epr}) as
\begin{equation}\label{epr1}
\sigma=2\pi\int_{\sigma_{0}} dE\sum_{j,k=1}^{N}f(x_{E,j})(x_{E,k}-x_{E,j})
|T_{k,j}(E)|^{2}.
\end{equation}
We shall say that the model has a nontrivial scattering in the channel $(j_{0},k_{0})$
if $ T_{j_{0},k_{0}}(E)\neq 0$ (on a subset of $\sigma_{0}$ of positive Lebesgue measure). As expected, the non-triviality of the scattering implies strict positivity of the entropy production:
\begin{corollary}
Suppose that the scattering is nontrivial in the channel  $(j_{0},k_{0})$ and in addition $(\beta_{j_{0}},\mu_{j_{0}})\neq (\beta_{k_{0}},\mu_{k_{0}})$
(i.e. at least $\beta_{j_{0}}\neq \beta_{k_{0}}$ or 
 $\mu_{j_{0}}\neq \mu_{k_{0}}$ ). Then
\begin{equation}\label{e>0}
\sigma >0.
\end{equation}
\end{corollary}
{\em Proof.} For the time reversal symmetric case  i.e.
$|T_{l,m}(E)|=|T_{m,l}(E)|$ (\ref{e>0}) follows from a simple  argument in \cite{AJPP}: by symmetrizing the sum in either (\ref{epr}) or (\ref{epr1}) one obtains:
\begin{equation}\label{epr2}
\sigma=\pi\int_{\sigma_{0}} dE\sum_{j,k=1}^{N}(f(x_{E,j})-f(x_{E,k}))(x_{E,k}-x_{E,j})
|T_{k,j}(E)|^{2}
\end{equation}
and then (\ref{e>0}) follows from the fact that $f(x)$ is strictly decreasing.
In the general case the argument is a bit more involved and, as expected, mimic the proof of Klein's inequality. Let $F(x)$ be a primitive of $f(x)$. Since $f(x)$ is strictly decreasing, $F(x)$ is strictly concave i.e 
\begin{equation}\label{conc}
F(x) \leq F(y)+f(y)(x-y)
\end{equation}
and the equality is reached only for $x=y$. From (\ref{conc}) and (\ref{normal1}) it follows that
\begin{eqnarray}\label{cF}
& \sum_{j=1,k}^{N}f(x_{E,j})(x_{E,k}-x_{E,j})
|T_{k,j}(E)|^{2}\geq \sum_{j,k=1}^{N}(F(x_{E,k})-F(x_{E,j})|T_{k,j}(E)|^{2}
\nonumber
\\
&
=\sum_{j,k=1}^{N}F(x_{E,k})(|T_{k,j}(E)|^{2}-|T_{j,k}(E)|^{2})=0.
\end{eqnarray}
and the equality is attained only if for all pairs $(j,k)$ and all (a.e) $E$, 
$(x_{E,k}-x_{E,j})|T_{k,j}(E)|=0$. Then the observation that $x_{E,k_{0}}-x_{E,j_{0}}=0 $ holds true for at most one energy finishes the proof .

{\bf Acknowledgments}

 I would like  to thank Claude-Alain Pillet 
 for 
stimulating discussions about NESS theory in general and about the
derivation of the Landauer-B\"uttiker
formula in particular.
This research has been supported  by the CEEX Grant
05-D11-45/2005.  This note 
 has been partly  written  during a visit at Aalborg
University. Both the financial support and the hospitality of the
Department of Mathematical Sciences, Aalborg University, are
gratefully acknowledged.


\begin{thebibliography}{99}

\bibitem{AJS}
W. O. Amrein, J. M. Jauch, K. B. Sinha: {\em Scattering Theory in
  Quantum Mechanics}, Benjamin, 1977.

\bibitem{AJPP}
W. Aschbacher, V. Jaksic, Y. Pautrat, C.-A. Pillet: Topics in
non-equilibrium statistical mechanics. mp$\_$arc 05-207. 

\bibitem{AJPP1}
W. Aschbacher, V. Jaksic, Y. Pautrat, C.-A. Pillet:Transport properties of ideal Fermi gases (in preparation).

\bibitem{AS}
W.Aschbacher, H. Spohn: A remark on the strict positivity of the entropy production. Lett. Math. Phys. {bf 75}, 17-23 (2006).

\bibitem{AEGSS}
J. E. Avron, A. Elgart, G-M. Graf, L. Sadun, K. Schnee: Adiabatic
charge pumping in open quantum systems. Commun. Pure. Appl. Math. {\bf
  57}, 528-561 (2004).

\bibitem{BBMH}
A. H. Barnett, M. Blaauboer, A. Mody, E. J. Heller: Mesoscopic
scattering in the half plane: Squeezing conductance through a small
hole. Phys. Rev. B
{\bf 63}, 245312 (2001).


\bibitem{BS}
H.U. Baranger, A.D. Stone: Electrical linear response theory in an
arbitrary magnetic field: A new Fermi-surface formation. Phys. Rev. B
{\bf 40}, 8169-8193 (1989).

\bibitem{B}
M. B\"uttiker: Four terminal phase-coherent
conductance. Phys.Rev. Lett. {\bf 57}, 1761-1764 (1986).

\bibitem{BILP}
M. B\"uttiker, Y. Imry, R. Landauer,
S. Pinhas: Generalized many-channel conductance with application to
small rings.  Phys. Rev. B
{\bf 31} 6207-6215 (1985).

\bibitem{CJM}
H. D.  Cornean, A. Jensen, V. Moldoveanu:  A rigorous proof of the
Landauer-B\"uttiker formula. 
 J. Math. Phys. {\bf 46}, 042106  (2005). 

\bibitem{FL}
D. S. Fisher, P.A. Lee: Relation between conductivity and transmission
matrix.
Phys. Rev. B {\bf 23}, 6851-6854 (1981).

\bibitem{FMU}
J. Fr\"{o}lich, M. Merkli, D. Ueltschi: Dissipative transport: Thermal contacts and tunnelling junctions. Ann. Henri Poincar\'{e} {\bf 4}, 897-945 (2003).

\bibitem{GW}
M. L. Goldberger, K. M.   Watson: Collision Theory,  Snd Edition, Dover (2004).

\bibitem{HDQEB}
A. W. Holleitner, C. R. Decker, H. Qin, K. Eberl, R. H. Blick:
Coherent coupling of two quantum dots embedded in an Aharonov-Bohm
interferometer.
Phys. Rev. Lett. {\bf 87 }, 256802 (2001).

\bibitem{IL}
Y. Imry, R. Landauer: Conductance viewed as
transmission. Rev. Mod. Phys.
{\bf 71}, S306-S312 (1999).

\bibitem{JKP}
V. Jaksic, E. Kritchevski, C.-A. Pillet: Mathematical theory of the
Wigner-Weisskopf Atom. mp$\_$arc 05-333 (To appear in Lecture Notes in
Mathematics, 2006).

\bibitem{JP}
V. Jaksic, C.-A. Pillet: On entropy production in quantum statistical mechanics. Comm. Math. Phys. {\bf 217}, 285-293 (2001).

\bibitem{MPD}  
P. Mavropoulos, N. Papanikolaou, P. H. Dederichs:
Korringa-Kohn-Rostoker formalism for ballistic transport.
Phys. Rev. B {\bf 69}, 125104 (2004).

\bibitem{MMS}
M. Merkli, M. M\"{u}ck, I. M. Sigal Theory of non-equilibrium stationary states as a theory of resonances. Existence and properties of NESS. arXiv: math-ph 0603006.

\bibitem{N}
G. Nenciu: A general proof of Landauer-B\"{u}ttiker formula. arXiv: math-ph 0603030.

\bibitem{RS}
M. Reed and B. Simon: {\em Methods of Modern Mathematical Physics.
Vol. III. Scattering Theory}, Academic Press, 1979.

\bibitem{Ru1}
D. Ruelle: Natural non-equilibrium states in quantum statistical mechanics. J. Stat. Phys. {\bf 98}, 57-75 (2000).

\bibitem{Ru2}
D. Ruelle: Entropy production in quantum spin systems. Comm. Math.Phys. {\bf 224}, 3-16 (2001).

\bibitem{Y1}
D. Yafaev: {\em Scattering Theory: Some Old and New Problems}, LNM
{\bf 1735}, Springer, 2000.

\bibitem{Y2}
D. Yafaev: {\em Mathematical Scattering Theory},  AMS, 1992.


\end{thebibliography}
\end{document}